%% file: ms.tex
\documentclass[sigconf,nonacm]{acmart}
\input{packages}
\input{macros}

\AtBeginDocument{%
  \providecommand\BibTeX{{%
    \normalfont B\kern-0.5em{\scshape i\kern-0.25em b}\kern-0.8em\TeX}}}

\begin{document}

\title{Complexity Results on Register Pushdown Automata}

\author{Ryoma Senda}
\email{ryoma.private@sqlab.jp}
\orcid{0000-0003-3596-4427}
\affiliation{%
  \institution{Graduate School of Informatics, Nagoya University}
  \streetaddress{Furo-cho, Chikusa}
  \city{Nagoya}
  \country{Japan}
  \postcode{464-8601}
}

\author{Yoshiaki Takata}
\email{takata.yoshiaki@kochi-tech.ac.jp}
\orcid{0000-0003-4982-0264}
\affiliation{%
  \institution{Graduate School of Engineering, Kochi University of Technology}
  \streetaddress{Tosayamada, Kami City}
  \city{Kochi}
  \country{Japan}
  \postcode{782-8502}
}

\author{Hiroyuki Seki}
\email{seki@i.nagoya-u.ac.jp}
\orcid{0000-0003-2001-7507}
\affiliation{%
  \institution{Graduate School of Informatics, Nagoya University}
  \streetaddress{Furo-cho, Chikusa}
  \city{Nagoya}
  \country{Japan}
  \postcode{464-8601}
}

\renewcommand{\shortauthors}{Senda, et al.}

\begin{abstract}
Register pushdown automata (RPDA) is
an extension of classical pushdown automata to handle data values in a restricted way.
RPDA attracts attention
as a model of a query language for structured documents with data values.
The membership and emptiness problems for RPDA are known to be EXPTIME-complete.
This paper shows the membership problem becomes PSPACE-complete and NP-complete
for nondecreasing and growing RPDA, respectively, while the emptiness
problem remains EXPTIME-complete for these subclasses.
\end{abstract}




\maketitle

\input{introduction}
\input{rpda}
\input{conclusion}


\bibliographystyle{ACM-Reference-Format}
\bibliography{main}

\input{references}



\end{document}

%% file: packages.tex
\usepackage{adjustbox}
\usepackage[linesnumbered,ruled,vlined]{algorithm2e}
\usepackage{booktabs}
\usepackage{amsmath}
\usepackage{amssymb}
\usepackage{bmpsize} 
\usepackage{booktabs}
\usepackage[shortlabels]{enumitem}
\usepackage{latexsym}
\usepackage{xspace}
\usepackage{here}

%% file: macros.tex
\newtheorem{example}{Example}

\newtheorem{theorem}{Theorem}

\newtheorem{corollary}{Corollary}

\newcommand{\Nat}{\mathbb{N}}
\newcommand{\Natz}{\mathbb{N}_0}

\renewcommand{\epsilon}{\varepsilon}

\newcommand{\done}{\Rightarrow}
\newcommand{\dast}{\stackrel{\ast}{\Rightarrow}}

\makeatletter
\newcommand{\xRightarrow}[2][]{%
\ext@arrow 0055{\Rightarrowfill@}{#1}{#2}%
}
\def\Rightarrowfill@{\arrowfill@\Relbar\Relbar\Rightarrow}
\newcommand{\xLeftarrow}[2][]{%
\ext@arrow 0055{\Leftarrowfill@}{#1}{#2}%
}
\def\Leftarrowfill@{\arrowfill@\Leftarrow\Relbar\Relbar}
\newcommand{\xLongleftrightarrow}[2][]{%
\ext@arrow 0055{\llrafill@}{#1}{#2}%
}
\def\llrafill@{\arrowfill@\Leftarrow\Relbar\Rightarrow}
\makeatother

%% file: introduction.tex
\section{Introduction}
%
There are many computational models having mild power of processing data values, 
including extensions of finite automata \cite{Bo02,KF94,LTV15,NSV04}, 
first-order and monadic second-order logics with data equality \cite{BDMSS09,BDMSS11} and 
linear temporal logics with freeze quantifier \cite{DLN07,DL09}.
Among them, 
register automata (abbreviated as RA) \cite{KF94} is a natural extension of finite automata
defined by incorporating registers 
as well as the equality test between an input data value and the data value kept in a register. 
Recently, attention has been paid to RA  
as a computational model of a query language
for structured documents such as XML because 
a query on a document can be specified as 
the combination of a regular pattern and 
a condition on data values \cite{LV12,LMV16}.
For query processing and optimization, 
the decidability (hopefully in polynomial time) 
of basic properties of queries is desirable. 
The most basic problem is
the membership problem that asks for a given query $q$ 
and an element $e$ in a document whether $e$ is in the answer set of $q$. 
The satisfiability or (non)emptiness problem asking whether the answer set of a given query 
is nonempty is also important because if the answer set is empty, 
the query can be regarded as redundant or meaningless when query optimization is performed.
The membership and emptiness problems for RA were 
 shown to be decidable \cite{KF94} and 
their computational complexities were also analyzed \cite{DL09,SI00}. 

While RA have the power of expressing regular patterns on 
{\em paths} in a document, 
it cannot represent 
patterns over branching paths that can be represented by 
some query languages such as XPath. 
Register context-free grammars (RCFG) and register pushdown automata (RPDA) 
were proposed in \cite{CK98}
as extensions of classical context-free grammars (CFG) and pushdown automata (PDA), 
respectively, in a similar way to extending FA to RA. 
Note that there are related but different extensions, 
tree automata with data, to deal with patterns over branching paths and data values
\cite{Fi10,Fi12,FS11,JK07,KT08,Se06}.
In \cite{CK98}, properties of RCFG and RPDA were shown including 
the equivalence in language representation power of RCFG and RPDA, 
the decidability of the membership and emptiness problems, 
and the closure properties. 
In \cite{STS18}, 
the computational complexity of the above decision problems for RCFG was investigated. 
In \cite{STS19}, the complexity of these problems for RPDA was also investigated but 
the analysis for subclasses of RPDA was not enough. 

In this paper, 
we discuss the computational complexity of 
the decision problems for some subclasses of RPDA. 
A $k$-RPDA has a finite-state control, $k$ registers and a pushdown stack (or stack in short) where
each cell of the stack can store a data value. 
A transition of $k$-RPDA is either a pop, replace or push transition. 
We introduce subclasses of RPDA called non-decreasing RPDA and growing RPDA
and investigate the computational complexity of the membership and emptiness problems 
for these subclasses. 

The main results of \cite{STS18,STS19} and this paper are summarized in Table \ref{tb:results}. The results for non-decreasing RPDA and growing RPDA are the contribution of this paper.
Note that the complexity of the membership problems is
in terms of both the size of a grammar or an automaton and the size of an input word
({\em combined complexity}). 
The complexity of the membership problem on the size of an input word only ({\em data complexity}) 
is in P for general RCFG and RPDA \cite{STS18}. 
It is desirable that the data complexity is small 
while the combined complexity 
is rather a criterion of the expressive succinctness of the query language.

\begin{table}[htb]
\caption{Complexity results on RCFG and RPDA}
\begin{center}
\small
\begin{tabular}{|c||c|c|c|} \hline
& \shortstack{general\\RCFG} 
& \shortstack{$\varepsilon$-rule free\\RCFG}
& \shortstack{growing\\RCFG}
\\ \cline{2-4}
& \shortstack{general\\RPDA} 
& \shortstack{non-decreasing\\RPDA}
& \shortstack{growing\\RPDA}
\\ \hline
Membership & EXPTIMEc & PSPACEc  & NPc      \\ \hline
Emptiness  & EXPTIMEc & EXPTIMEc & EXPTIMEc \\ \hline
\end{tabular}
\smallskip\par
`c' stands for `complete.'\\
The results for RCFG and their subclasses were given in \cite{STS18}\\
and those for general RPDA were given in \cite{STS19}. 
\end{center}
\label{tb:results}
\end{table}

%% file: rpda.tex
\section{Definitions}

A register pushdown automaton (abbreviated as RPDA) was originally defined in \cite{CK98}
as a pushdown automaton with registers over an infinite alphabet
and the equivalence between RCFG and RPDA was shown in \cite{CK98}.
In this section, we define RPDA as a pushdown automaton over the product of a finite alphabet
and an infinite set of data values, following recent notions \cite{LV12,LTV15}.
Note that these differences are not essential.

\subsection{Preliminaries}
Let $\Nat=\{1,2,\ldots\}$ and $\Natz=\{0\} \cup \Nat$.
We assume an infinite set $D$ of data values as well as
a finite alphabet $\Sigma$.
For a given $k\in\Natz$ specifying the number of registers,
a mapping $\theta: [k] \to D$ is called an assignment
(of data values to $k$ registers) where $[k]=\{1,2,\ldots,k\}$.
We assume that a data value $\bot\in D$ is designated as the initial value of a register and the initial bottom symbol in a stack. We denote by $\theta_\bot$ the register assignment that assigns
the initial value $\bot$ to every register.
Let $\Theta_k$ denote the class of assignments to $k$ registers.
For $\theta,\theta'\in \Theta_k$, we write $\theta' = \theta[i\leftarrow d]$
if $\theta'(i)=d$ and $\theta'(j)=\theta(j)$ ($j\not=i$).

Let $F_k$ denote the set of guard expressions over $k$ registers
defined by the following syntax rules:
\begin{quote}
\[ \psi :=
 \mbox{tt} \mid x_i^{=} \mid x_{top}^{=} \mid \psi \vee \psi \mid \neg \psi
\]
where $x_i \in \{x_1, \ldots, x_k\}$.
\end{quote}
For $d, e\in D$ and $\theta\in \Theta_k$, the satisfaction of $\psi \in F_k$ by $(\theta,d, e)$ is
recursively defined as follows.
Intuitively, $d$ is a current data value in the input,
$e$ is the data value at the stack top,
$\theta$ is a current register assignment,
$\theta,d,e\models x_i^{=}$ means that the data value assigned to the $i$-th register by $\theta$
is equal to the input data value $d$ and $\theta, d, e\models x_{top}^=$ means that the input data value $d$  equals to the data value $e$ at the stack top.
\begin{itemize}
\item $\theta,d,e \models x_i^{=}$ iff $\theta(i)=d$
\item $\theta,d,e \models x_{top}^{=}$ iff $d=e$
\end{itemize}
The other cases are defined in an ordinary way. In addition, let $\mbox{ff}=\neg \mbox{tt}, \psi_1 \wedge \psi_2 = \neg(\neg \psi_1 \vee \neg \psi_2), x_i^{\not=}=\neg(x_i^=)$ and $x_{top}^{\not=}=\neg(x_{top}^=)$ for $\psi_1, \psi_2\in F_k$ and $i\in[k]$.
\par
For a finite alphabet $\Sigma$ and a set $D$ of data values disjoint from $\Sigma$,
a {\em data word} over $\Sigma\times D$ is a finite sequence of elements of $\Sigma\times D$
and a subset of $(\Sigma \times D)^{\ast}$ is called a {\em data language} over $\Sigma\times D$.
$|\beta|$ denotes the cardinality of $\beta$ if $\beta$ is a set
and the length of $\beta$ if $\beta$ is a finite sequence.

\subsection{Register pushdown automata}

\label{def:rpda}
For $k\in\Natz$, a $k$-{register pushdown automaton} ($k$-RPDA) over a finite alphabet $\Sigma$ and
a set $D$ of data values is a tuple $\mathcal{A}=(Q, q_0, \delta)$ where
$Q$ is a finite set of states, $q_0 \in Q$ is the initial state
and $\delta$ is a finite set of transition rules having one of the following forms:
\begin{itemize}
\item $(p, \psi, i) \stackrel{a}{\rightarrow} (q, \varepsilon)$  (or $(p, \psi) \stackrel{a}{\rightarrow} (q, \varepsilon)$) (pop rule)
\item $(p, \psi, i) \stackrel{a}{\rightarrow} (q, j_1)$  (or $(p, \psi) \stackrel{a}{\rightarrow} (q, j_1)$) (replace rule)
\item $(p, \psi, i) \stackrel{a}{\rightarrow} (q, j_1 j_2)$  (or $(p, \psi) \stackrel{a}{\rightarrow} (q, j_1 j_2)$) (push rule)
\end{itemize}
where $p, q\in Q$, $a\in(\Sigma\cup\{\varepsilon\})$, $i, j_1, j_2 \in[k]$, and $\psi\in F_k$. A rule is called an $\varepsilon$-rule if $a=\varepsilon$.

$D$ is used as a stack alphabet.
For a state $q\in Q$, a register assignment $\theta\in \Theta_k$, a data word $w \in (\Sigma\times D)^*$,
and a stack $u \in D^*$, $(q, \theta, w, u)$ is called
an {\em instanteneous description (abbreviated as ID)} of $k$-RPDA $\mathcal{A}$ and $|u|$ is the stack height of this ID.
For two IDs $(q, \theta, w, u), (q', \theta', w', u')$, we say that $(q, \theta, w, u)$ can transit to $(q', \theta', w', u')$, written as $(q, \theta, w, u) \done_\mathcal{A} (q', \theta', w', u')$ if there exists a rule $(p, \psi, i) \stackrel{a}{\rightarrow} (q', J) \in \delta$ (resp.  $(p, \psi) \stackrel{a}{\rightarrow} (q', J)$), data values $d, e\in D$ and $u''\in D^{\ast}$ such that
\begin{align*}
\theta,d,e \models\psi, ~~\theta'=\theta[i\leftarrow d] \text{ (resp. $\theta'=\theta$)},& \\
\begin{aligned}
w &= \begin{cases}
    (a, d)w' & a\in \Sigma, \text{ or}\\
    w' & a=\varepsilon
  \end{cases}
\end{aligned}
,  u=eu'', &\text{ and}\\
\begin{aligned}
u' &= \begin{cases}
    u'' & J=\varepsilon, \\
    \theta'(j_1) u'' & J=j_1, \text{ or} \\
    \theta'(j_1) \theta'(j_2) u'' & J=j_1 j_2.
  \end{cases}
\end{aligned}
\end{align*}
For an $\varepsilon$-rule, we can choose an arbitrary data value $d$ that satisfies $\theta,d,e \models\psi$ and $\theta'=\theta[i\leftarrow d]$.
\par
Let $\dast_\mathcal{A}$ be the reflexive transitive closure of $\Rightarrow_\mathcal{A}$. We abbreviate $\done_\mathcal{A}$ and $\dast_\mathcal{A}$ as $\done$ and $\dast$ if $\mathcal{A}$ is clear from the context.

For a $k$-RPDA $\mathcal{A}=(Q, q_0, \delta)$ and $w\in(\Sigma\times D)^*$,
if $(q_0, \theta_\bot, w, \bot) \dast (q, \theta, \varepsilon, \varepsilon)$
for some $q\in Q$ and $\theta\in \Theta_k$,
then we say $\mathcal{A}$ accepts $w$,
$(q_0, \theta_\bot, w, \bot) \dast (q, \theta, \varepsilon, \varepsilon)$
is an {\em accepting run} of $w$ in $\mathcal{A}$, and the number of the transitions in the run
is called the {\em length} of the run.
Let $L(\mathcal{A})=\{w\in(\Sigma\times D)^* \mid \mathcal{A}\text{ accepts }w\}$,
which is the language recognized by $\mathcal{A}$.

We say that an RPDA $\mathcal{A}$ is {\em non-decreasing} if every $\varepsilon$-rule of $\mathcal{A}$ is either a replace rule or a push rule.
We say that an RPDA $\mathcal{A}$ is {\em growing} if every $\varepsilon$-rule of $\mathcal{A}$ is a push rule.
\noindent
\begin{example}
Let $\mathcal{A}_1=(Q, q_0, \delta)$ be the $2$-RPDA, where
\begin{itemize}
\item $Q=\{q_0, q_1, q_2, q_3\}$, and
\item $\delta=\{(q_0, \mbox{tt}, 1)\stackrel{a}{\to}(q_1, 11),
(q_1, x_1^{\not=}, 2)\stackrel{b}{\to}(q_1, 22),
(q_1, \mbox{tt})\stackrel{\varepsilon}{\to}(q_2, \varepsilon),
(q_2, x_{top}^{=}\wedge x_1^\neq)\stackrel{b}{\to}(q_2, \varepsilon),
(q_2, x_{top}^{=})\stackrel{a}{\to}(q_3, \varepsilon)\}$.
\end{itemize}
In a run of $\mathcal{A}_1$, the first input data value is pushed and loaded to the first register in $q_0$, and then an arbitrary number of input data values different from the first register are pushed in $q_1$. After the current state is nondeterministically changed to $q_2$ by the $\varepsilon$-rule (the third rule in $\delta$), the data values in the stack are popped and compared with the remaining input data values. Hence, $L(\mathcal{A}_1)=\{(a, d_0)\cdots(b, d_n)$\\$(b, d_n)\cdots(a, d_0)\mid d_0\neq d_i \text{ for } i\in[n], n\geq 0\}$.
\begin{figure}[htbp]
\label{fig:2.1}
  \centering
  \includegraphics[width=6cm]{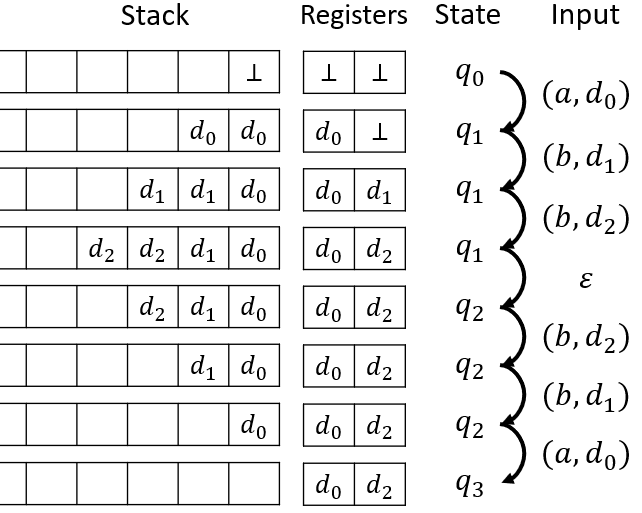}
  \caption{The transitions in the accepting run of $(a,d_0)(b,d_1)(b, d_2)(b, d_2)(b,d_1)(a,d_0)$ in $\mathcal{A}_1$ ($d_i\neq d_0$ for $i\neq 0$).}
\end{figure}
\end{example}

\subsection{Turing machine}
In this section, we define {\em Turing machine} (abbreviated as TM) and some notions for TM. We use TM for proving Theorem \ref{the:2}.
\par
A TM is a tuple
$M = (Q, \Gamma, \Sigma, \delta, q_{0}, F)$ where
\begin{itemize}
      \item $Q$ is a finite set of states,
      \item $\Gamma$ is a finite set of tape symbols containing a special symbol representing
      {\em blank}, $\sqcup \in \Gamma \backslash \Sigma$,
      \item $\Sigma \subseteq \Gamma \backslash \{\sqcup\}$ is a set of input symbols,
      \item $\delta: Q \times \Gamma \rightarrow Q \times \Gamma \times \{L, R\}$
      is a state transition function,
	\item $q_0 \in Q$ is the initial state, and
      \item $F\subseteq Q$ is the set of accepting states.
\end{itemize}
\par
For a state $q \in Q$, a tape content $\alpha \in \Gamma^{*}$
and a head position $j\in[|\alpha|]$,
$(q, \alpha, j)$ is called ID of $M$.
For two IDs $(q, \alpha, j), (q', \alpha', j')$,
we say that $(q, \alpha, j)$ can transit to $(q', \alpha', j')$
, written as
$(q, \alpha, j) \to (q', \alpha', j')$ if
\begin{align*}
\begin{aligned}
 \exists a,b\in\Gamma,\ \exists\beta, \gamma &\in\Gamma^{*},\ \exists m\in\{L, R\} \text{ such that}\\
 \ |\beta|=j-1,\ \alpha=&\beta a\gamma, \delta(q, a)=(q', b, m)\\
\alpha'&=
\begin{cases}
    \rlap{$\beta b\sqcup$}\hphantom{j-1} & \text{if $|\alpha| = j$ (i.e.\ $\gamma=\varepsilon$) and $j' = j + 1$},\\
    \beta b\gamma & \text{otherwise},
    \end{cases}
\\
j' &= \begin{cases}
    \mbox{max}(1, j - 1) & m=L,\ \text{or}\\
    j + 1 & m=R.
  \end{cases}
\end{aligned}
\end{align*}
Let $\xrightarrow{*}$ be the reflexive transitive closure of $\to$.
\par
For a TM $M = (Q, \Gamma, \Sigma, \delta, q_{0}, F)$
and $u \in \Sigma^{*}$,
if $(q_{0}, u, 1) \xrightarrow{*} (q_f, u', j)$ for some $q_f\in F, u'\in\Gamma^*$ and $j\in \Nat$, then $M$ accepts $u$.
Let $L(M) = \{u \in \Sigma^{*} \mid \text{$M$ accepts $u$}\}$, which is
the language recognized by $M$.
For a function $f: \Natz \to \Natz$, 
if for any $u\in \Sigma^{*}$ and
any $(q, \alpha, j)$ such that $(q_0, u, 1) \stackrel{*}{\to} (q, \alpha, j)$,
$|\alpha| \le f(|u|)$ holds, then
we say that $M$ is an $f(n)$-space bounded TM\@.
If $M$ is a $p(n)$-space bounded TM for a polynomial $p(n)$,
$M$ is a polynomial space bounded TM\@.

\section{Computational Complexity}
In \cite{STS18}, it was shown that the membership and emptiness problems for RCFG are EXPTIME-complete. Those problems for RPDA were also shown to be EXPTIME-complete by bidirectional poly-time equivalence transformations between RCFG and RPDA in \cite{STS19}.
\begin{theorem}[\cite{STS19}]
\label{the:8.1}
The membership and emptiness problems for RPDA are EXPTIME-complete.
\end{theorem}
\par
In this section, we first show that the membership problems for non-decreasing and growing RPDA are PSPACE-complete and NP-complete, respectively. Although the lower-bound proofs are similar to those for $\varepsilon$-rule free and growing RCFG in \cite{STS18}, we present formal description of the reductions in these proofs to make the paper self-contained.

\begin{theorem}
\label{the:2}
The membership problem for non-decreasing RPDA is PSPACE-complete.
\end{theorem}
{\bf Proof} \quad
If a given RPDA is non-decreasing, we can decrease the length of a stack only by using non-$\varepsilon$-pop rules with reading an input data word. Therefore, for an input data word $w$, the height of every stack appearing in an accepting run of $w$ is at most $|w|$ because every accepting run must finish with empty stack. Hence, the membership problem is in PSPACE.

To prove PSPACE-hardness, we use a poly-time reduction from the membership problem for polynomial space bounded TM. In the reduction, we simulate tape contents of a given TM $M$
by a register assignment of the RPDA $\mathcal{A}$ constructed from $M$.
\par
Assume that we are given a $p(n)$-space bounded TM
$M = (Q_M, \Gamma, \allowbreak \Sigma, \delta_M,\allowbreak q_{0}, F)$
where $p(n)$ is a polynomial
 and an input $u \in \Sigma^{\ast}$ to $M$.
 Then, we construct $(|\Gamma| + p(|u|))$-RPDA $\mathcal{A}= (Q_\mathcal{A}, T_{(1,0)}, \delta_\mathcal{A})$ over a singleton alphabet $\{a\}$ and an arbitrary set $D$ of data values that satisfies
 $u \in L(M) \Leftrightarrow (a, \bot) \in L(\mathcal{A})$, where
\begin{align*}
Q_\mathcal{A} &=
\begin{aligned}[t]
& \{T_{(i,j)} \mid 1 \leq j < i \leq |\Gamma|\} \cup \{ T_{(1,0)} \}\\
& {}\cup \{W_{i} \mid 0 \leq i \leq |u|\}\\
& {}\cup \{A_{q}^{(i, j)}
        \mid q \in Q_M,\ i\in [|\Gamma|],\ j\in [p(|u|)]\} \\
& {}\cup \{B_{q}^{(i, j)}
        \mid q \in Q_M,\ i\in [|\Gamma|],\ j\in [p(|u|)]\}\\
& {}\cup\{E\}
\end{aligned}
\end{align*}
and $\delta_\mathcal{A}$ is constructed as follows.
Without loss of generality,
we assume that
$\Gamma=\{1,2,\ldots,|\Gamma|\} \subseteq \Nat$ 
and $1$ is the blank symbol of $M$.
In the following,
we denote the $i$th element of a sequence $\alpha$ by $\alpha_i$
(i.e., $\alpha=\alpha_1\alpha_2\ldots\alpha_{|\alpha|}$).

\begin{itemize}
\item
We construct transition rules that load different data values in the first $|\Gamma|$ registers.
Note that we keep the initial value $\bot$ in the first register.
To the $i$th register ($2\leq i \leq |\Gamma|$), a data value different from $\bot$ is
assigned by Rule~(\ref{eq:1a}), and
that data value is guaranteed to be different from the value of every
$j$th register ($2\le j<i$) by Rule~(\ref{eq:1b}).
\begin{align}\textstyle
(T_{(i-1,i-2)}, x_{1}^{\ne}, i) & \stackrel{\varepsilon}{\to} (T_{(i,1)}, 1)
  \quad \text{for $2 \le i \le |\Gamma|$}, \label{eq:1a} \\
(T_{(i,j-1)}, x_{i}^{=}\land x_{j}^{\ne}) & \stackrel{\varepsilon}{\to} (T_{(i,j)}, 1)
  \quad \text{for $2 \le j < i \le |\Gamma|$}. \label{eq:1b}
\end{align}
%

\item
To express the initial tape contents $u$, we construct the following transition rules
that load data values corresponding to the symbols in $u$ from left to right
into $(|\Gamma|+1)$th to $(|\Gamma|+|u|)$th registers:
\begin{align}
(T_{(|\Gamma|,|\Gamma|-1)}, \mbox{tt}) & \stackrel{\varepsilon}{\to} (W_{0}, 1),\\
(W_{i-1}, x_{u_{i}}^{=}, |\Gamma| + i) & \stackrel{\varepsilon}{\to} (W_{i}, 1)
\quad \text{for $i\in [|u|]$}.
\end{align}
\item Let $s(m) =
-1$ if $m = L$ and $s(m)=
1$ if $m = R$.
For encoding the state transition and accepting condition of $M$ by $\mathcal{A}$,
we introduce a state $A_{q}^{(i,j)}$
for $q\in Q_M$, $i\in [|\Gamma|]$, and $j\in [p(n)]$.
$A_{q}^{(i,j)}$ represents a part of an ID $(q, \alpha, j)$ of $M$
where $i=\alpha_j$, i.e.~the tape symbol at the head position.
The remaining information about $\alpha$ of $(q,\alpha,j)$ will be represented by
a register assignment of $\mathcal{A}$.
More precisely,
the content of $(|\Gamma|+j)$th register (i.e.~$\theta(|\Gamma|+j)$)
equals the data value $\theta(\alpha_j)$ representing the tape symbol $\alpha_j$ for $j\in [|\alpha|]$
and $\theta(|\Gamma|+j) = \bot$ for $|\alpha|<j\le p(|u|)$.
Let $\theta_{\alpha}$ denote such a register assignment that represents
the tape contents $\alpha$.
We illustrate the correspondence between
an ID of $M$ and a state and a register assignment of $\mathcal{A}$ in Fig.~\ref{fig:2}.
\begin{figure}[tb]
  \begin{center}
%
      \newlength{\Bwd}%
      \settowidth{\Bwd}{$\alpha_{|\alpha|}$}%
      \def\B#1{\makebox[\Bwd][c]{$#1$}\Bstrut}%
      \def\Bstrut{\mbox{\vrule width0pt height .7\Bwd depth .3\Bwd}}%
      \def\Bldots{\makebox[2\Bwd][c]{$\ldots$}}%
    \tabcolsep=1pt
	$A_q^{(\alpha_j,j)}$
    \begin{tabular}{|c|c|c|c|c|c|c|c|}
    \hline
    \qquad$\Gamma$\qquad\null
     & \B{\alpha_1} & \B{\alpha_2} & \Bldots & \B{\alpha_{|\alpha|}}
    & \B{\bot} & \Bldots & \B{\bot} \\
    \hline
    \end{tabular}
    \caption{$\mathcal{A}$'s state and registers that correspond to $M$'s ID $(q,\alpha,j)$.}
    \label{fig:2}
  \end{center}
\end{figure}
\begin{itemize}
\item To derive the states corresponding to the initial ID of $M$,
we construct the following rule:
\begin{align}
(W_{|u|}, \mbox{tt}) &\stackrel{\varepsilon}{\to} (A_{q_{0}}^{(u_{1}, 1)}, 1).
\end{align}
\item
Consider $A_{q}^{(i, j)}$ and let
$\delta_M(q, i)=(q', b', m')$.
For each $a \in \Gamma$,
we construct the following rules:
\begin{align}
(A_{q}^{(i, j)}, x_{b'}^{=}, |\Gamma| + j) & \stackrel{\varepsilon}{\to} (B_{q'}^{(a, \mbox{max}(1, j+s(m')))}, 1).
\label{eq:5}
\end{align}
We also construct the following rule
for each $q'\in Q_M$, $a\in\Gamma$, and $j'\in [p(n)]$:
\begin{align}
(B_{q'}^{(a, j')}, x_{a}^{=} \land x_{|\Gamma| + j'}^{=})
&\stackrel{\varepsilon}{\to} (A_{q'}^{(a, j')}, 1).
\label{eq:6}
\end{align}
\end{itemize}
\item
Finally, we construct for each $q_f\in F$ the following rules to express accepting IDs:
\begin{align}
(A_{q_{f}}^{(i, j)}, x_1^=) &\stackrel{a}{\to} (E, \varepsilon).
\label{eq:10}
\end{align}
\end{itemize}
\par\noindent
We can show for each ID $(q, \alpha, j)$,
\begin{eqnarray}
(q, \alpha, j)\xrightarrow{*}(q_f, u', j') \text{ for some } q_f\in F, u'\in\Gamma^* \text{ and } j'\in \Nat \nonumber\\
\text{iff}\
(A_{q}^{(\alpha_j, j)}, \theta_{\alpha}, (a, \bot), \bot) \xRightarrow{*}_\mathcal{A} (E, \theta, \varepsilon, \varepsilon) \text{ for some } \theta\in\Theta_k
\label{the:1:property:1}
\end{eqnarray}
by induction on the length of the run of $M$
for only if part and
by induction on the length of the run of $\mathcal{A}$ for if part.

We can easily prove that
$(T_{(1,0)},\theta_\bot, (a, \bot), \bot) \xRightarrow{*}_\mathcal{A}$\\$ (A_{q_0}^{(u_1,1)},\theta_u, (a, \bot), \bot)$,
and moreover,
if $(T_{(1,0)},\theta_\bot, (a, \bot), \bot) \xRightarrow{*}_\mathcal{A}\allowbreak (E, \theta, \varepsilon, \varepsilon)$ for some $\theta\in\Theta_k$, then
this run must be\\
$(T_{(1,0)},\theta_\bot, (a, \bot), \bot)\allowbreak \xRightarrow{*}_\mathcal{A}
 (A_{q_0}^{(u_1,1)},\theta_u, (a, \bot), \bot) \xRightarrow{*}_\mathcal{A} (E, \theta, \varepsilon, \varepsilon)$.
\\By letting $(q,\alpha,j)=(q_0,u,1)$ in property~(\ref{the:1:property:1})
and by the above-mentioned fact,
we obtain $u \in L(M) \Leftrightarrow ((T_{(1,0)},\theta_\bot, (a, \bot), \bot) \xRightarrow{*}_{\mathcal{A}} (E, \theta, \varepsilon, \varepsilon)\allowbreak \text{ for some }\allowbreak \theta\in\Theta_k) \Leftrightarrow (a, \bot) \in L(\mathcal{A})$.

\begin{theorem}
\label{the:3}
The membership problem for growing RPDA is NP-complete.
\end{theorem}
{\bf Proof} \quad
If a given RPDA is growing, for an input data word $w$, an accepting run of $w$ applies $\varepsilon$-push rules at most $|w|$ times. Therefore, the length of an accepting run does not exceed $2|w|+1$. Hence, the membership problem is in NP.

We prove NP-hardness by a polynomial-time reduction from the satisfiability problem for 3-Conjunctive Normal Form (3CNF).
Let $\phi = (a_1 \lor b_1 \lor c_1)\dots(a_m \lor b_m \lor c_m)$ be
a 3CNF over Boolean variables $y_1, \ldots, y_n$
where
each $a_i, b_i, c_i$ ($i\in[m]$) is a literal $y_j$ or $\overline{y_j}$
for some $j$ ($j\in[n]$).
For $i$ ($i\in[m]$), we define register number
$r_{a_i}$ as
$r_{a_i} = 2j$     if $a_i = y_j$ and
$r_{a_i} = 2j + 1$ if $a_i = \overline{y_j}$.
We also define the same notation $r_{b_i}$ and $r_{c_i}$
for $b_i$ and $c_i$.
We construct the growing $(2n+1)$-RPDA $\mathcal{A} = (Q, q_0, \delta)$
over $\Sigma = \{a\}$
where $Q = \{q_0, P_0,\dots,P_n, C_0, \dots, C_m, E\}$ and
\begin{align*}
\delta ={} & \{(q_0, \text{tt}, 1) \stackrel{a}{\to} (P_0, 1)\} \\
& {}\cup\{(P_{i-1}, x_{1}^{=}, 2i+j) \stackrel{a}{\to} (P_{i}, 1) \mid i\in[n],\ j\in\{0,1\} \}\\
& {}\cup\{(P_n, x_1^=) \stackrel{a}{\to} (C_0,1)\}\\\
& {}\cup\{(C_{i-1},
        x_r^=)
        \stackrel{a}{\to} (C_i, 1) \mid i\in[m],\
        r\in\{ r_{a_i},r_{b_i},r_{c_i} \} \} \\
& {}\cup\{(C_m, x_1^=) \stackrel{a}{\to} (E, \varepsilon) \}.
\end{align*}
%
The first register of the constructed RPDA $\mathcal{A}$
is used for keeping a data value (possibly) different from $\bot$,
and we use that value and $\bot$ for representing tt and ff, respectively.
$\mathcal{A}$ nondeterministically loads the value representing tt
to exactly one of the $(2i)$th and $(2i+1)$th registers for each $i$,
to encode a truth value assignment
to $y_1, \overline{y_1}, y_2, \overline{y_2}, \ldots, y_n, \overline{y_n}$.
Then $\mathcal{A}$ reads the value of one of the literals $a_i, b_i, c_i$
for each clause $a_i\lor b_i\lor c_i$ in $\phi$.
It is not difficult to show that $\phi$ is satisfiable if and only if $(a, d)^{n+m+3} \in L(\mathcal{A})$,
where $d$ is an arbitrary data value in $D\setminus\{\bot\}$.
Since $(a,d_1)^{n+m+3}\in L(\mathcal{A})$ iff $(a,d_2)^{n+m+3}\in L(\mathcal{A})$ for
any $d_1,d_2\in D\setminus\{\bot\}$,
we can choose any $d\in D\setminus\{\bot\}$ to make
the input data word for the membership problem.
Hence, we have shown NP-hardness of the problem. \\\par
For both of the RPDA constructed in the proofs of Theorem \ref{the:2} and \ref{the:3}, the height of the stack appearing in any accepting run is at most one. This fact implies the following property.
\begin{corollary}
The membership problems for RA with and without $\varepsilon$-transition are PSPACE-complete and NP-complete, respectively.
\end{corollary}
Note that NP-completeness of the membership for RA without $\varepsilon$-transition was first proved in \cite{SI00}.

\begin{theorem}
The emptiness problems for non-decreasing RPDA and growing RPDA are both EXPTIME-complete.
\end{theorem}
{\bf Proof} \quad
By theorem \ref{the:8.1}, it suffices to show that the problem is EXPTIME-hard for growing RPDA. We give a poly-time reduction from the emptiness problem for general RPDA.
From an arbitrary RPDA $\mathcal{A}=(Q,q_0,\delta)$, we construct the growing RPDA $\mathcal{A}_g=(Q,q_0,\delta_g)$, where
\begin{quote}
$\delta_g=\{(q, \psi, i)\stackrel{a}{\rightarrow} (q',J)\mid (q, \psi, i)\stackrel{a}{\rightarrow} (q',J)\in \delta\text{ or }(q, \psi, i)\stackrel{\varepsilon}{\rightarrow} (q',J)\in \delta\}$.
\end{quote}
For this growing RPDA $\mathcal{A}_g$, obviously $L(\mathcal{A})=\emptyset\Leftrightarrow L(\mathcal{A}_g)=\emptyset$.  Hence, the emptiness problem for growing RPDA is EXPTIME-hard.

%% file: conclusion.tex
\section{Conclusion}
We have discussed the computational complexity of
the membership and emptiness problems for RPDA.
The combined complexity of the membership problem for general RPDA is
EXPTIME-complete and decreases when we consider subclasses of RPDA
while the data complexity is in P for general RPDA.
The emptiness problem for RPDA remains EXPTIME-complete even if we
restrict RPDA to be growing. It is an interesting problem left as future study to investigate
whether the inclusions among the classes of languages of general, non-decreasing and growing RPDA are proper or not.

\par
Introducing a logic such as $\mbox{FO}(\sim)$, $\mbox{EMSO}(\sim)$ and
LTL$\downarrow$ on data trees that corresponds to or subsumes
RPDA is a future study.
Also, introducing recursive queries such as datalog in relational databases
and fixed point logics as related logical foundations \cite{AHV95,Li04}
would be an interesting topic to be pursued.
\par
This work was supported by JSPS KAKENHI Grant Number\\
JP19H04083. 